 \documentclass[final,5p,times,twocolumn]{elsarticle}

\usepackage{amssymb}
\begin{document}

\begin{frontmatter}



\title{Towards compact Free Electron Laser based on laser plasma accelerators}


\author{Marie Emmanuelle Couprie}

\address{Synchrotron SOLEIL, L'Orme des Merisiers, Saint-Aubin, 91 192 Gif-sur-Yvette, France}

\begin{abstract}
The laser invention more than fifty years ago was a major scientific revolution. 
Among the different possible gain media, the Free Electron Lasers (FEL) uses free electrons in the period permanent magnetic field of an undulator, covering wavelengths from far infra red to X-ray, with easy tuneability and high peak power. Nowadays, the advent of tuneable intense (mJ level) short pulse FELs with record peak power (GW level) in the X-ray domain sets a major step in laser development, and enables to explore new scientific areas, such as deciphering molecular reactions in real time, understanding functions of proteins. 
Besides, lasers have also been considered for driving plasma electron acceleration. A high-power femtosecond laser is focused into a gas target and resonantly drives a nonlinear plasma wave in which plasma electrons are trapped and accelerated with high energy gain of GeV/m. Nowadays, laser wakefield acceleration became reality and electron beams with multi-GeV energies, hundreds pC charge, sub-percent energy spread and sub-milliradian divergence can be produced. It is relevant to consider a FEL application to quality these laser plasma produced electrons. After having described the FEL panorama, the strategies towards laser plasma based acceleration based FELs will be discussed, including the mitigation of the large energy spread and divergence of these beams should be mitigated.

\end{abstract}


\end{frontmatter}



\section{Introduction : the origins of the Free Electron Laser}
\label{Introduction}

\subsection{Stimulated emission}

In 1927, Einstein (1879-1955, Nobel prize in 1921) predicted the energy enhancement by atom desexcitation \cite{einstein1917quantentheorie} in the analysis of the black-body radiation, while absorption and spontaneous emission were the known light matter interactions at that time. This process was named in 1924 stimulated emission \cite{VanVleck24, VanVleck_RevModPhys77}. First, a photon is absorbed and drives an atom to an excited state. The excited atom being unstable, it emits a spontaneous photon after a duration depending on the lifetime of the excited level. Besides, when a photon is absorbed by an excited atom, two photons with identical wavelength, direction, phase, polarization are emitted, while the atom returns to its fundamental state. This stimulated emission emission was seen as addition of photons to already existing photons, and not as the amplification of a monochromatic wave with conservation of its phase, without the notion of light coherence. 

\subsection{Vacuum tubes}

\begin{figure}
\centering\includegraphics[width=.7\linewidth]{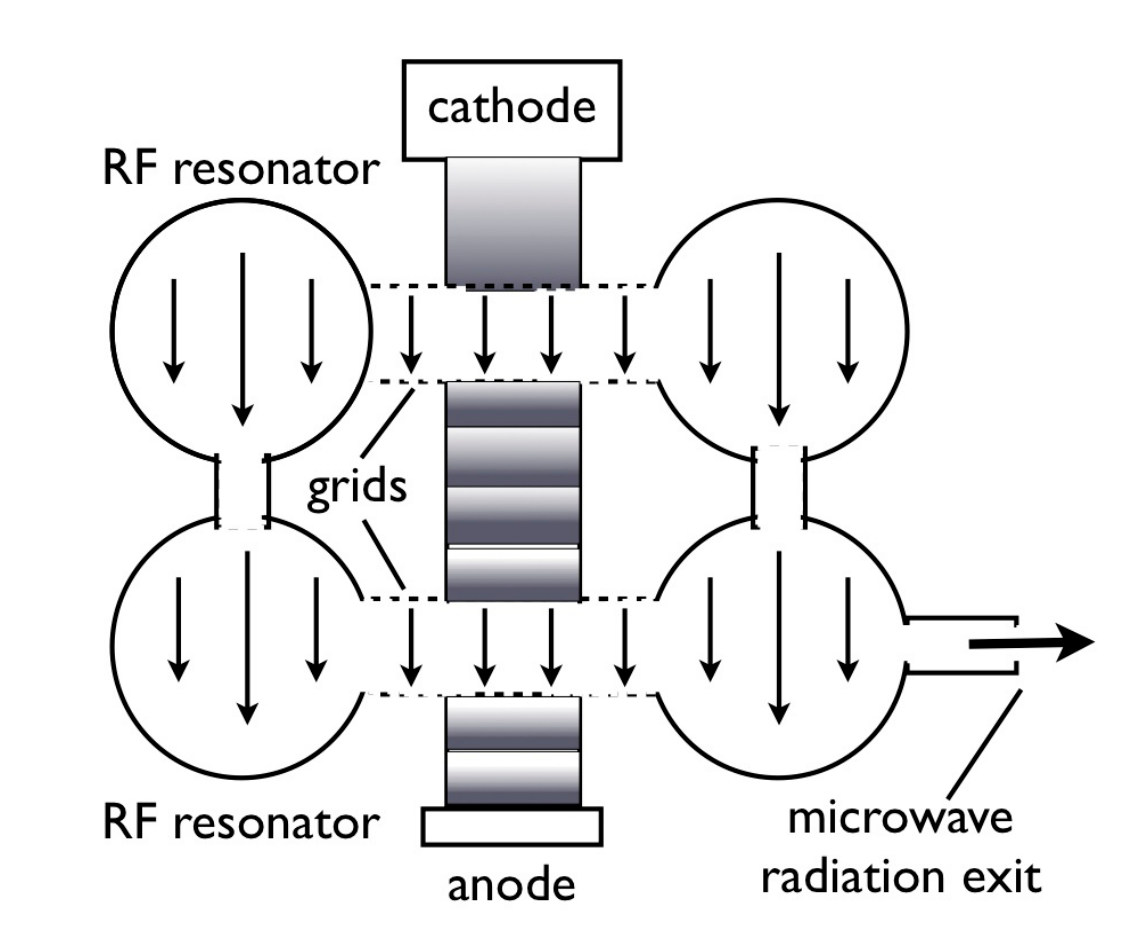}
\caption{Klystron principle : klystron scheme (electron bunching by energy modulation in the klystron drift space, electron accumulation in bunches, and RF field amplification due to phased electron in the second klystron cavity}
\label{fig:Fig_klystron_scheme}
\end{figure}

The electron beam in vacuum tubes witnessed a rapid and spectacular development in the beginning of the twentieth century for the current amplifier applications such as radiodiffusion, radar detection, where high frequency oscillations were needed. In vacuum tubes, a free electron of relativistic factor $\gamma$ given by $ \gamma=\frac{{E}}{m_{o} c^2}$ (with $E$ its energy,  $m_{o}$ the electron mass,  $e$ the particle charge and $c$ the speed of light) interacts with an electromagnetic wave of electric field $\vec{E}$ :  $\vec{E}=\vec{E} \sin{(ks-\omega . t)}$ with $k$ the wave number and $\omega$ the pulsation according to $\frac{ d\gamma}{dt} = \frac{e\vec{\beta}.{\vec{E}}}{mc}$ with $\beta$ the normalised electron velocity. 
In a klystron \cite{Varian39} (see Fig.~\ref{fig:Fig_klystron_scheme}), electrons generated on a cathode enter in a first metallic cavity where an input GHz electric field is applied (an electric field oscillating at a frequency $\nu=2 \pi f $ of several  $GHz$). The sign of their energy gain $\Delta W_{1}  \simeq e c E. \Delta s .  \cos{( \omega t)} $ depends on the moment t when they arrive inside the cavity, $\Delta W_{1}$ is thus modulated in time at a temporal period $T = \frac{ 2 \pi} {  \omega}  $ or spatial period   $ 2 \pi c \beta / \omega  $. $<\Delta W_{1}>_{electrons}=0$ since the electrons have different phases. Then, the electrons enter into the drift space and accumulate in bunches. In the second cavity of interaction region $L_{2}$, the bunched electrons have the same phase with respect to the electromagnetic wave. The second energy exchange $\Delta W_{2} = %\sum_{electrons}  \int_{0}^{l_{2}} e c \vec{\beta}. \vec{E_{RF}} dt = %
N_{e}e c E L_{2}   \cos{( \omega t)} $ 
leads to very high electric field gains, from 3 to 6 orders of magnitude. 

%Consider then a  relativistic electron of energy $E$ and velocity $v$ with respect to the laboratory frame. Its relativistic factor $\gamma$ is given by  $ \gamma=\frac{{E}}{m_{o} c^2}$ (with  $m_{o}$ the electron mass,  $e$ the particle charge and $c$ the speed of light). It can be expressed as : $\gamma = \frac{1}{\sqrt {1-\beta ^2}} $ with  $\beta$ the normalized velocity of the electron, expressed as : $\beta=\frac{v}{c}$. 

\subsection{Synchrotron radiation}

Synchrotron radiation, the electromagnetic radiation emitted by accelerated charged particles, is generally produced artificially in particle accelerators. Its theoretical foundations established at the end of the nineteenth century \cite{Larmor1897, Lienard1898} were developed further \cite{Schott1907, Ivanenko, Oliphant, McMillan, Veksler, Schwinger1946, Schwinger1949}. After the first particle energy loss on a $100~MeV$ betatron \cite{Blewett1946}, the first synchrotron radiation was observed in the visible tangent to the electron orbit one year later on the $70~MeV$ General Electric synchrotron, of $29.3~m$ radius and $0.8~T$ peak magnetic field \cite{Elder1947obsRS}.  Radiation is emitted in a narrow cone of aperture $1/\gamma$.

%\subsection{The undulator}
Radiation emitted by relativistic electrons performing transverse oscillations was first considered \cite{Ginzburg1947}.
The electromagnetic field created by a relativistic particle in a periodic permanent magnetic field (produced by undulators, consisting of a succession of alternated poles)  \cite{Motz1951, motz1959radiation}  %(see Fig.~\ref{fig:schemeundHalbach.pdf}) 
was calculated and observed \cite{Motz1953, Combes1955}. For a planar undulator generating a sinusoidal magnetic field 
$\vec{B_{u}}=B_{u} \cos \left(\frac{2\pi}{\lambda_{u}}s\right) \vec{z} = B_{u} \cos ( k_{u} s) \vec{z}$ with the undulator wavenumber $k_{u}$ : $k_{u}=\frac{2\pi}{\lambda_{u}}$, the emitted radiation along the undulator periods interfere constructively according to $\lambda_{n}  =  \frac{\lambda_{u}}{2n\gamma^{2} } (1+  K_{u}^{2}/2+ \gamma^2\theta^2)$ with the deflection parameter $K_{u}= \frac{eB_{u}\lambda_{u}}{2\pi m_{o} c}$ and $\theta$ the observation angle. The radiation is tuneable by changing the magnetic field or the electron energy. 

% of period $\lambda_u$
%\begin{figure}[!h]
%\centering\includegraphics[width=.4\linewidth]{schemeundHalbach.jpg}
%\caption{Planar undulator scheme, creating a periodic vertical magnetic field (green) with two arrays of permanent magnets, electron trajectory (blue). }
%\label{fig:schemeundHalbach.pdf}
%\end{figure}

\subsection{The laser and maser discoveries}

For doing a «quantum» microwave source using stimulated emission in molecules instead of the amplification by an electron beam, an excited molecule is introduced in a microwave cavity resonant at the frequency of the molecule transition. In 1954, the first MASER  (Microwave Amplification by Stimulated Emission of Radiation) is operated  in the micro-waves  \cite{Townes54} at Columbia Univ with $NH_3$ molecule. 
For reaching the optical spectral range, an open Fabry-Perot type resonant cavity  (In a cavity in which the light makes round trips between the two mirrors on which it is reflected) \cite{Schawlow58} is used instead of a resonant cavity on its fundamental mode that would becoming extremely small ($\sim$ 1 $\mu m$). These "optical lasers" were named LASER for Light Amplification by Stimulated Emission of Radiation \cite{gould1959laser}. "For wavelengths much shorter than those of the ultraviolet region, maser-type amplification appears to be quite impractical.  Although using of a multimode cavity is suggested, a single mode may be selected by making only the end walls highly reflecting, and defining a suitably small angular aperture. Then extremely monochromatic and coherent light is produced \cite{Schawlow58}."  Lasers were achieved experimentally, with  Ruby \cite{Maiman60, maiman1961stimulated}, He--Ne  \cite{javan1961population}, AsGa  \cite{hall1962coherent} and others \cite{keyes1962recombination}. Limits in extending lasers towards very short wavelengths were underlined : ``As one attempts to  extend maser operation towards very short wavelengths, a number of new aspects and problems arise, which require a quantitative reorientation of theoretical discussions and considerable modification of the experimental techniques used'' ; ``These figures show that maser systems can be expected to operate successfully in the infrared, optical, and perhaps in the ultraviolet regions, but that, unless some radically new approach is found, they cannot be pushed to wavelengths much shorter than those in the ultraviolet region''  \cite{Schawlow58}.

\subsection{The free electron laser invention}

J. M. J. Madey (1943--2016) considered that ``A. Schawlow and C. Townes descriptions of masers and lasers coupled with the new understanding of the Gaussian eigenmodes of free space offered a new approach to high frequency operation that was not constrained by the established limits to the capabilities of electron tubes'' \cite{Madey14} and  examined whether there was ``a Free Electron Radiation Mechanism that Could Fulfill these Conditions''  \cite{Madey15}, considering different possible radiation processes. Stimulated Compton Scattering appeared very promising  \cite{dreicer1964kinetic}. Already investigated earlier  \cite{Pantell68, marcuse1962stimulated, landecker1952possibility, twiss1958radiation}
The Compton backscattering (CBS) radiation resulting from a head-on collision between a laser pulse and a bunch of relativistic particles has a high energy  $E_{\mathrm{CBS}}$ : $E_{\mathrm{CBS}} =   \frac{4 \gamma^{2} E_{\mathrm{ph}}}{1 +(\gamma\theta)^2 }$ with $E_{\mathrm{ph}} $ the energy of the initial photon beam, $\theta$ the angle between the CBS photons and the electron beam trajectory. The CBS radiation could be tuneable by a change of the energy of the relativistic electrons. The scattered radiation presents a low divergence for relativistic electron beams (i.e. $\gamma \gg 1$), which radiation cone is reduced to $1/\gamma$. 
J. M. J. Madey had the idea to make the phenomenon more efficient by using the magnetic field of an undulator  \cite{Madey71}: ``Relativistic electrons can also not tell the difference between real and virtual incident photons, permitting the substitution of a strong, periodic transverse magnetic field for the usual counter-propagating real photon beam'' \cite{Madey15}. 

\begin{figure}[!h]
\centering\includegraphics[width=.5\linewidth]{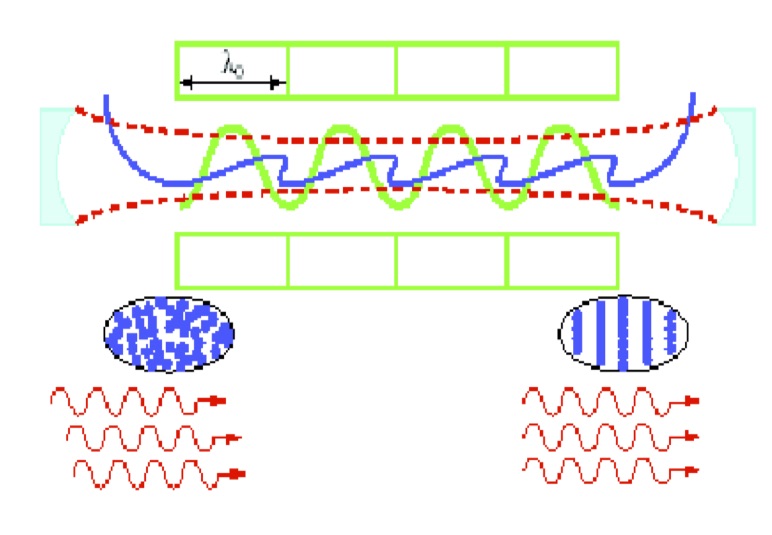}
\caption{Scheme of the FEL oscillator with the gain medium consisting of relativistic electrons in the undulator}
\label{fig:Fig_FELoscillator_simple.pdf}
\end{figure}

His proposed scheme (see Fig.~\ref{fig:Fig_FELoscillator_simple.pdf}) includes thus the electron beam in the undulator field as the gain medium, and the optical resonator, as for lasers. After first gain calculations in quantum mechanics \cite{Madey71}, theory was developed in with various approaches : plasma \cite{sprangle1974stimulated}, distribution functions \cite{hopf1976classical},  relativistic motion of the electrons in the undulator and energy exchange  \cite{colson1976theory}. 
The electron beam progressing in the undulator emits synchrotron radiation, which is stored in the optical resonator. An energy exchange between the optical wave and the electrons takes place, leading the microbunching of the electron bunch at $\lambda_n$ separation (electron position being depending of the energy). The electrons are thus set in phase, radiate coherently and the light is amplified. Saturation takes place by enhancement of energy spread, or the undulator resonance condition unsatisfied.  

\section{Free Electron Laser development}
\label{Section 1}

\subsection{Free electron lasers oscillators and harmonic generation}

\begin{figure}[!h]
\centering\includegraphics[width=.6\linewidth]{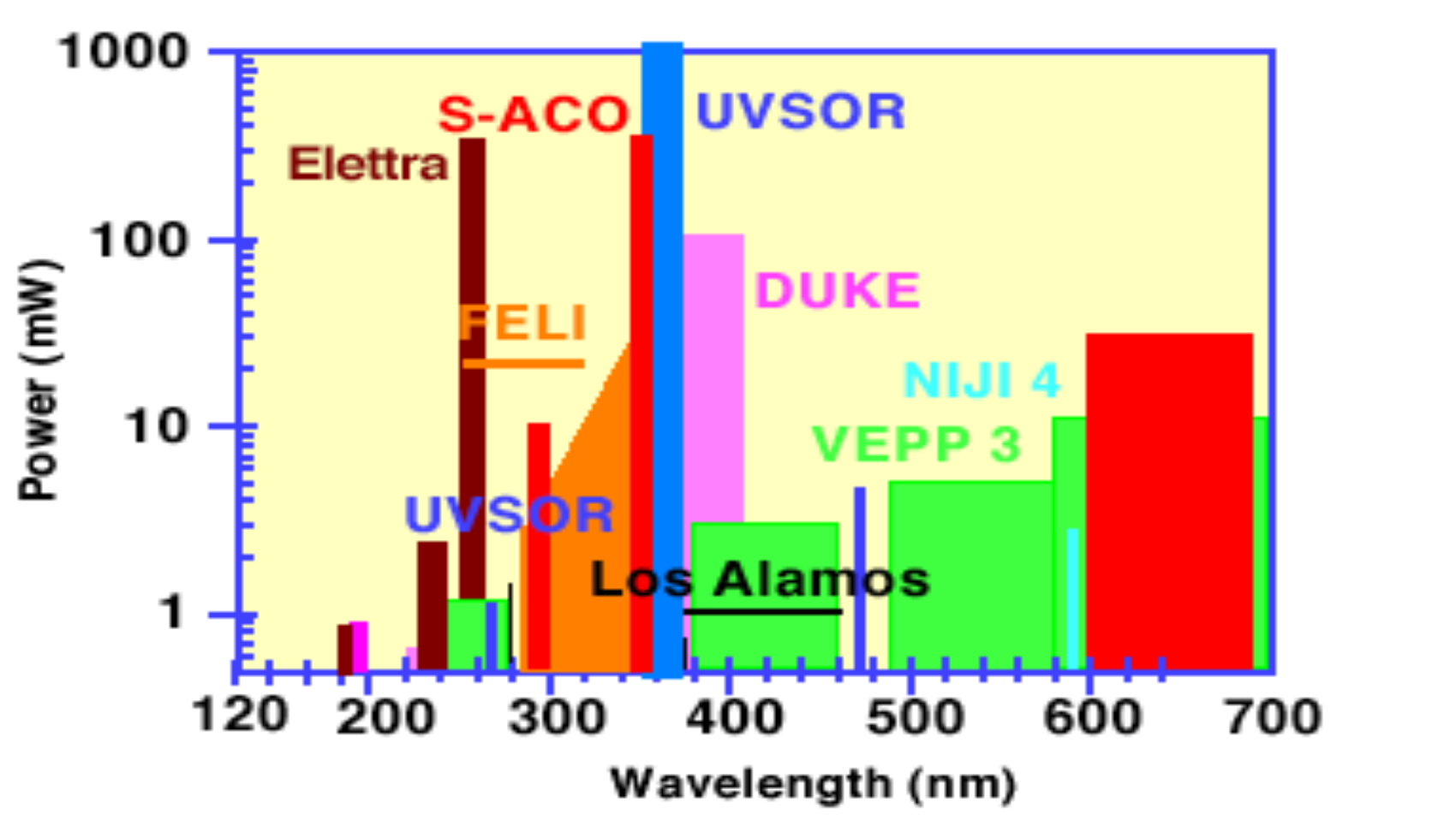}
\caption{Short wavelength FEL oscillators}
\label{fig:Fig_shortwavelengthFELoscillator.pdf}
\end{figure}

The first experimental demonstration of the FEL amplification (single pass gain of $7\%$) in the infra-red was performed in Stanford in 1976 \cite{Elias76} on the super-conducting linear accelerator. The first FEL oscillation was achieved at 3.4 $\mu$m in 1977 360~mW average power, corresponding to  an estimated 7 (500 intracavity)~kW peak power \cite{Deacon77}. The second worldwide FEL oscillation was then obtained in 1983 on the ACO storage ring, in the visible  \cite{Billardon83}, and then followed by Coherent harmonic generation  \cite{Girard84, prazeres1987first} in the UV and VUV usng a Nd--Yag laser. The Stanford FEL has been operated with a tapered undulator in order to enhance the efficiency \cite{edighoffer1984variable}, enabling the output power to be enlarged  \cite{Kroll80taper}. FEL oscillation was obtained at Los Alamos in 1983 at  $9-11~ \mu m$, with nine orders of magnitude of power growth and a net gain of $17~ \%$, leading to an intra-cavity peak power of  $20~ MW$  \cite{warren1983}. A $4 \%$ efficiency \cite{feldman1989high} was reached with undulator tapering. 
Besides linear accelerators and storage rings \cite{couprie1997storage, couprie2000short}, FEL oscillators were then developed on various types of accelerators, such as Van de Graff, microtrons, energy recovery linacs. The developed FEL oscillators enabled to cover from infra-red to VUV spectral range (Fig.~\ref{fig:Fig_shortwavelengthFELoscillator.pdf}) with the shorter wavelength obtained on the ELETTRA storage ring FEL \cite{Marsielttra2002, trovo2002operation}. The limit is set somehow by the gain value compared to the mirror losses  \cite{gatto2002high} submitted to drastic irradiation conditions \cite{garzella1995mirror}. FEL oscillators present a very high degree of coherence, both in transverse thanks to the optical resonator and in longitudinal close to the Fourier limit thanks to multi-passes. 

\subsection{Single  pass high gain regime FEL}

\begin{figure}[!h]
\centering\includegraphics[width=.6\linewidth]{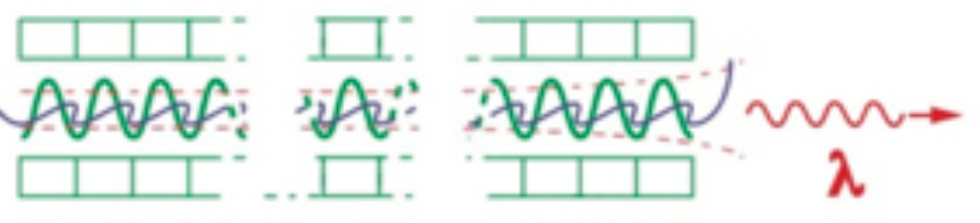}
\caption{FEL Self Amplified Spontaneous Emission (SASE) configuration : spontaneous emission emitted along the undulator amplified in one single pass.}
\label{fig:Fig_SASEconfig.pdf}
\end{figure}

In parallel to high gain FEL studies \cite{kroll1978stimulated, sprangle1975stimulated, hopf1976strong, haus1981noise, dattoli1981progress}, the production of coherent radiation from a self-instability, without the use of an optical resonator was considered \cite{Kondratenko79, Kondratenko79-48, Kondratenko80} and even a short wavelengths \cite{derbenev1982possibility}. The system starts from noise with the undulator spontaneous emission which is amplified it in the high gain regime until saturation (see Fig.~\ref{fig:Fig_SASEconfig.pdf}). More precisely,  the electrons communicate with each other through the radiation and the space charge field; they "self bunch" on the scale of the radiation wavelength periods. A  collective instability occurs where the electrons have nearly the same phase and emit collectively coherent synchrotron radiation \cite{bonifacio1982cooperative, Bonifacio84} over a cooperative length. After a "lethargy" period required for the initial pulse to build up, the light is then amplified exponentially with a gain length $L_{go} = \frac{1}{\sqrt{3} 2 k_u \rho_{FEL}} (1 + (\sigma_{\gamma}/\rho_{FEL})^{2}$ with $\rho_{FEL} $ the so-called Pierce parameter and $\sigma_{\gamma}$ the energy spread.  
This regime is called Self Amplified Spontaneous Emission (SASE), in reference to the Amplified Spontaneous Emission in conventional lasers. 
%$k_u$ the undulator wavenumber and
The SASE spectral bandwidth is given by the Pierce parameter $\frac{\Delta \lambda}{\lambda} = \rho_{FEL}$, and the saturation power by $P_{sat} = \rho_{FEL}E I_p$ with $I_p$ the peak current. Typically, the saturation power is reached after roughly $20$ gain lengths, at the saturation length $L_s$. 
%$E$ the electron beam energy, and 
The interaction between the electrons is only effective over a cooperation length, the slippage (slipping of the  emitted by one electron moves ahead by one wavelength per undulator period) in one gain length, as $  L_{coop}=\frac{\lambda}{2 \sqrt{3}\rho_{FEL}}$ \cite{Bonifacio90}. The uncorrelated trains of radiation, which result from the interaction of electrons progressing jointly with the previously emitted spontaneous radiation, lead to spiky longitudinal and temporal distributions and poor longitudinal coherence, apart from single spike operation for low charge  short bunch regime  \cite{Din2009, Saldin98}. 

SASE experimental results were first obtained at long wavelength the mid eighties  (saturated high gain amplification in the mm waves ($34.6~GHz$) in (Lawrence Livermore National Laboratory / Lawrence Berkeley Laboratory (USA) collaboration) \cite{orzechowski1985microwave}, superradiant emission at $640~\mu m$ at MIT (USA) \cite{kirkpatrick1989high}, observation of bunching at  $8~mm$ \cite{gardelle1997high} and SASE \cite{lefevre1999self})  at CESTA (France). 
Then, SASE was observed in the infra-red ($20~-~40~\mu m$ at ISIR (Japan) \cite{okuda1993self}, at SUNSHINE (USA)  \cite{bocek1996observation}, at CLIO (France) in the mid-infrared ($5-10~ \mu m$) \cite{prazeres1997observation}, at BNL (USA) at 1064 and 633 nm  \cite{BabzienPhysRevE.57.6093}, at Los Alamos (USA) at $15~\mu m$ \cite{nguyen1998self}). Then five orders of magnitude of amplification and saturation at $12 \mu m$ have been achieved  (UCLA, Los Alamos, Stanford, Kurchatov collaboration) on  the Advanced Free Electron Laser (AFEL) linac at the Los Alamos National Laboratory \cite{hogan1998measurements, hogan1998measurementsbis}. Saturation at $845~nm$ has been observed on the VISA (Visible to Infrared SASE Amplifier) experiment on the Accelerator Test Facility (ATF) at Brookhaven National Laboratory (USA) \cite{tremaine2002experimental, murokh2003properties}. 

The beginning the the twentieth century saw the advent of the saturated SASE in the visible and UV (530  and 385 nm) in 2000 at Argonne National Laboratory (USA)  \cite{milton2000observation, Milton2001} on the Low-Energy Undulator Test Line (LEUTL). In parallel, nonlinear harmonics at $423$ and $281 ~nm$ were observed using the VISA SASE FEL at saturation \cite{tremaine2002experimental}. In the same years, a major step was achieved with the observation of  SASE radiation in the VUV spectral range, with a  $233~MeV$ electron beam from the Tesla Test Facility (Germany) presently called FLASH  using a photo-injector and superconducting accelerator modules ($6\pi ~mm.mrad$ emittance, $0.4~kA$ peak current, $0.13~ \%$ relative energy spread), enabling a gain of 3000 at $109~nm$ \cite{AndruszkowPhysRevLett.85.3825} in 2000 and then saturation \cite{ayvazyan2002new} in 2001, i.e. twenty- five years after the FEL invention. Tunability in  $80~-~ 120 ~nm$ range was demonstrated, and a very high degree of photon beam transverse coherence was observed. With higher peak current, the GW level (close to $1 ~\mu J$ energy) had been reached in the $95-105~nm$ spectral range \cite{Ayvazyan2002PhysRevLett.88.104802}  with a gain length of $67~cm$. 
These results competed the shortest wavelength achieved on a FEL oscillators (on a storage ring), making a turning point in the choice of the type of the FEL accelerator driver and configuration. The path towards the X-ray domain with SASE radiation was paved with new achievements, such as the SASE radiation in the $60~-~40~nm$ spectral range with an energy of $30~mJ$ on SCSS Test Accelerator (Japan) \cite{Shintake08, shintake2009stable}, $6.5~ nm$ \cite{schreiber2008operation} and $4.1~ nm$ \cite{schreiber2011first}, i.e. in the water window on the fundamental at FLASH.

Then, a new area started with the advent of hard X-ray FELs, with first LCLS in Stanford (USA)  at $0.15~nm$, with saturation after 60 m of undulators \cite{Emma10}, followed by SACLA (Japan) in 2011 down to $0.08~ nm$ \cite{Ishikawa12}, then PAL FEL (Korea) in 2016 \cite{kang2017hard}, Swiss FEL (Switzerland) \cite{SwissFEL} and European X FEL (Germany) \cite{Weise2017} in 2017, while new projects are under development. Operation at short wavelengths requires high beam energies for reaching the resonant wavelength, and thus long undulators ($0.1-1~km$ for $0.1~nm$) and high electron beam density (small emittance and short bunches) for ensuring a sufficient gain.  The obtention of SASE radiation at shorter wavelength benefited largely from the improvements of the linac electron beam performance, thanks to the development of photo-injectors \cite{sheffield1988alamos, feldman1991experimental, hartman1994initial, travier1991rf} and more generally of the accelerator developments towards colliders. 

\begin{figure}[!h]
\centering\includegraphics[width=.6\linewidth]{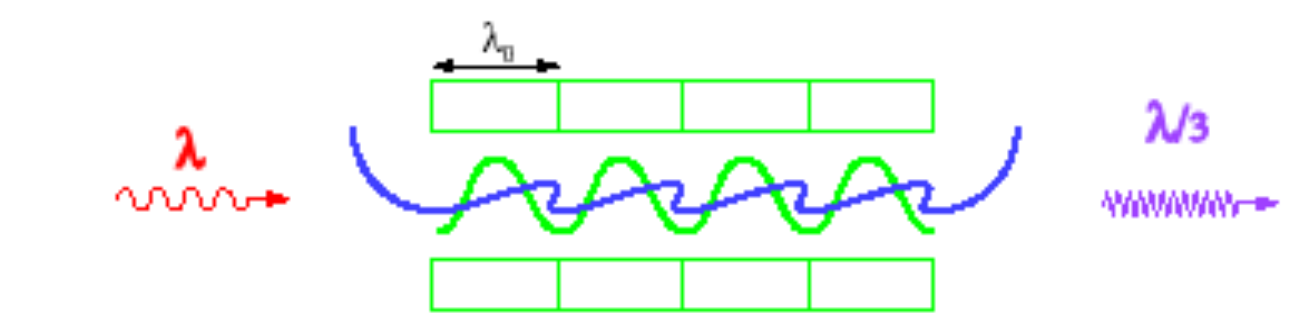}
\caption{Seeding scheme}
\label{fig:Fig_seeding}
\end{figure}
   
Besides the spectacular advent of the powerful tuneable FELs (mJ energy per pulse), the FEL pulse spiky spectral and temporal distributions with the associated jitter still provide some limitations for FEL use. Besides using low-charge short electron bunches \cite{reiche2008development}], a chirped electron bunch associated with an undulator taper \cite{GiannessiSASEchirpPhysRevLett.106.144801}, 
the longitudinal coherence of a single pass FEL can be significantly improved by seeding with an external laser spectrally tuned on the undulator fundamental radiation, while intensity fluctuations are reduced and saturation is reached earlier (see Fig.~\ref{fig:Fig_seeding}). 
Non linear harmonics can also be efficiently generated \cite{huang2000three, dattoli2005nonlinear, dattoli2007pulse} in different configurations such as the High Gain harmonic Generation  \cite {yu1991generation, yu2002theory, ben1992fresh} : A small energy modulation is imposed on the electron beam by its interaction with a seed laser in a first undulator (the modulator) tuned to the seed frequency, it is is then converted into a longitudinal density modulation thanks to a dispersive section (chicane) and in a second undulator (the radiator), which is tuned to the nth harmonic of the seed frequency, the micro-bunched electron beam emits coherent radiation at the harmonic frequency of the first one, which is then amplified in the radiator until saturation is reached \cite{Yu00HGHG}. High order harmonics generated in gas can also be used as a seed \cite{Lambert08}. Such a scheme can be put in cascade for wavelength reduction.
According to the seed characteristics with respect to that of the electron bunch, different regimes such as super radiance \cite{giannessi2005nonlinear}, pulse splitting  \cite{labat2009pulse, de2013chirped} can be observed. 

 FERMI@ELLETRA, the first seeded FEL users facility in Trieste (Italy) consists of two FEL branches, FEL 1 in the $100 ~- ~ 20~ nm$ via a single cascade harmonic generation, and FEL 2 in the $20 -  4~ nm$ via a double cascade harmonic generation \cite{allaria2010fermi, allaria2012highly, Allaria12NaturePhotonics}. Up-frequency conversion by a factor of 192 \cite{giannessi2014pulse}. The Dalian FEL (Dalian, China) covers $50 ~-  150 nm$ \cite{wang2017commissioning}. 
 Seeding with the FEL itself is also considered \cite{Feldhaus97, Geloni11} and is of particular interest  for the X-ray domain: a monochromator installed after the first undulator spectrally cleans the radiation before the last amplification in the final undulator. Recently, self-seeding  with the spectral cleaning of the SASE radiation has been experimentally demonstrated at LCLS \cite{Amann12, ratner2015experimental} and at SACLA \cite{Yabashi12}.

\begin{figure}[!h]
\centering\includegraphics[width=.8\linewidth]{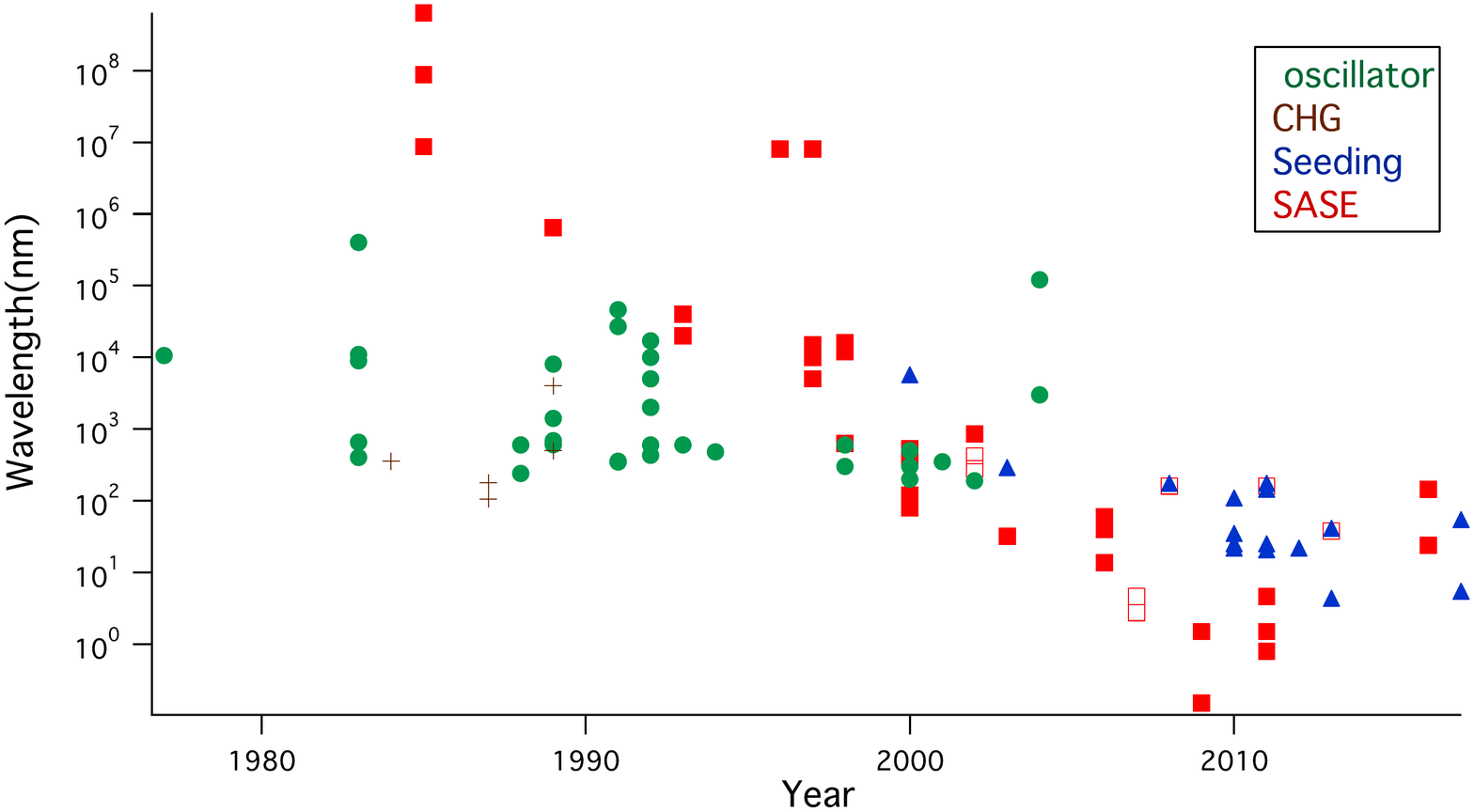}
\caption{Achieved FEL wavelengths versus year for various configurations (oscillators, coherent harmonic generation, SASE, seeding)}
\label{fig:Fig_Wavelengthvsyear.pdf}
\end{figure}

Fig.~\ref{fig:Fig_Wavelengthvsyear.pdf} shows the trend in FEL wavelength decrease versus years : up to the century change, FEL oscillators were the most suitable candidates, while afterwards, single pass FEL such as SASE with their improved versions in terms of temporal coherence (seeded FEL) appeared the most adequate. This turn is mainly due to the improvement of the linear accelerator technology, FEL community being benefiting from the developments of high brightness electron beams required for future linear colliders. The path has been long towards these unique tunable intense X-ray FELs, with some projects that did fail. More than forty years have been spend between the first FEL in the infra-red and the first X-ray FEL, both in Stanford.

Present developments are oriented in providing further advanced properties \cite{couprie2016strategies}.The two-colour FEL concept can be applied to the X-ray domain in the SASE regime, either tuning the two series of undulators at different wavelengths \cite{lutman2013experimental, marinelli2013multicolor, hara2013two}, the delay being adjusted by a chicane, or by using twin bunches at different energies \cite{marinelli2015high}, enabling also operation in the self-seeded case. In the external seeding case, one can take advantage of the pulse splitting effect \cite{labat2009pulse} combined with chirp \cite{de2013chirped, mahieu2013two}, or apply a double seeding \cite{allaria2013two, petralia2015two}. Several strategies are investigated in the quest towards in attosecond pulses and high peak power. The Echo Enabled Harmonic Generation \cite{Stupakov09} (EEHG) enables efficient up-frequency conversion by imprinting a sheet-like structure in phase space via a two successive electron-laser interactions in two undulators. Experimental results were obtained on harmonic 7  \cite{Xiang10}, 15 \cite{hemsing2014highly} and 75 \cite{hemsing2016echo} on the Next Linear Collider Test Accelerator and on the Shanghai FEL Test Facility \cite{Zhao12}. The trend is also to use superconducting high repetition rate linear accelerator for FEL line multiplexing and for preventing from space charge effects for some user experiments. 
Another approach investigates compactness besides seeding and up-frequency conversion, by using novel acceleration techniques, such as laser plasma acceleration \cite{Tajima}.

\section{Strategies towards LPA based FELs}
\label{LPA based FELs}

The laser invention led thus to free electron laser and to laser plasma acceleration. On can wonder then whether these two different paths could join again for developing a laser plasma acceleration based free electron laser. The idea arose ten years ago \cite{gruner2007design, Nakajima08}.
Issues related to this prospects are discussed.

\subsection{Performance of Laser Plasma Acceleration}

%\subsection{ Laser Plasma Acceleration}

Inspired by the laser invention, in parallel to the Free Electron Laser invention by J. M. J. Madey, emerged the idea of laser wavefield acceleration two years later. The concept has been described as follows by Tajima and Dawson  \cite{Tajima}: 
 "An intense electromagnetic pulse can create a wake of plasma oscillations through the action of the nonlinear ponderomotive force. Electrons trapped in the wake can be accelerated to high energy. Existing glass lasers of power density $10^{18}$ W/$cm^2$ shown on plasma densities of $10^{18}$ $cm^{-3}$  can yield GeV of electron energy per centimeter of acceleration distance. This acceleration mechanism is demonstrated through computer simulation. Applications to accelerators and pulsers are examined".  Indeed, an intense laser pulse drives plasma density wakes to produce, by charge separation, strong longitudinal electric fields, with accelerating gradient than can reach a 100 GV/m \cite{esarey2009physics, malka2008principles}, in which the electron with a proper phase can be efficiently accelerated. 
Following the high power laser development thanks to chirped pulse amplification \cite{strickland1985compression}, significant electron beam acceleration was achieved \cite{Geddes04, faure:Nat04, mangles:Nat04, Malka05, Leemans06}.   
 
 LWFA can nowadays produce electron beams in the few hundreds of MeV to severals GeV with a typical current of a few kiloamperes with reasonable beam characteristics (relative energy spread of the order of $1\%$, and a normalized emittance of the order of $ 1\pi$ mm\,mrad) 
 \cite{Faure06, Lu07, Cipiccia, Lundh, Rechatin, geddes:PRL2008, Kalmykov:PRL2009, faure:POP2010, Pollock:PRL2011, corde:NatComm2013, leemans:PRL2014, weingartner:PRSTAB2012, thaury:SciRep2015}. 
 However, all these "best" performance are usually not achieved simultaneously and depend on the chosen configuration  (bubble  \cite{esarey2009physics},  colliding scheme \cite{esarey:PRL1997}, optical transverse injection \cite{lehe2013optical}, shock front injection \cite{buck2013shock}, ionization injection \cite{clayton2010self, McGuffey:PRL2010}, plasma channel \cite{Geddes04}, frequency chirp \cite{pathak2012effect} ...) and on staging \cite{steinke:nature2016, hosokai2010electron}.

\subsection{A first step : observation of undulator radiation}

LWFA based undulator radiation has been observed, even at short wavelengths \cite{Schlenvoigt08, Fuchs09, Anania09, Lambert12} and recently at LUX at 9 nm \cite{MaierCFEL}. The quality of the spectra do not meet yet what is currently achieved and used on synchrotron radiation based facilities in terms of spectral bandwidth, intensity and stability.

\subsection{Conditions for SASE amplification}

Several conditions are required for the FEL amplification to be possible, and they set specifications on the electron beam. The Pierce parameter $\rho_{FEL}$ is expressed as  \cite{Huang07, pellegrini2016physics}: 

 \begin{equation}
\rho_{FEL} = \Big[ \frac{K_u [JJ] \omega_p}{4 \omega_u} \Big]^{2/3} = \frac{1}{2 \gamma k_u} \Big( \frac{\mu_o e^2 K_{u}^{2} [JJ]^2 k_u n_e }{4 m_o}\Big)^{1/3}
\end{equation}

with $\omega_p$ and $\omega_u$ the plasma and undulator pulsations, $\mu_o$ the magnetic permeability,  $[JJ]$ the planar undulator Bessel function difference term \cite{couprie2013synchrotron}. 
%$e$ the electron charge, , $m_o$ the electron mass
For the  energy modulation and bunching to be maintain for insuring sufficient gain,  the electron beam  should be rather "cold", its energy spread $\sigma_{\gamma}$ should be smaller than the bandwidth, i. e.: $\sigma_{\gamma} < \rho_{FEL}$.
  There should be a proper transverse matching (size, divergence) between the electron beam and the photon beam along the undulator for insuring the interaction. In consequence, the emittance $\epsilon_{n}$ should not be too large at short wavelength.  The FEL gain increases with the beam current provided that: $\frac{\epsilon_{n}}{\gamma} < \frac{\lambda}{4 \pi}$.  The radiation diffraction losses should be smaller than the FEL gain, i.e. the Rayleigh length should be larger than the gain length ($Z_{r} > L_{go}$). The condition can be smoothened in case of gain guiding. For long undulators, intermediate focusing is then put between undulator segments.
 High power short wavelength FELs require thus low emittance electron beams (much smaller than $ 100 ~ \pi mm.mrad$ and peak currents of the order of $100~ A$.

 \subsection{Issues related to LPA regarding FEL amplification}
Following undulator spontaneous emission observation, this new accelerating concept could thus be qualified by a FEL application.
But achieving it remains to be demonstrated: the difficulty comes from the intrinsic properties of the electron beam. Indeed, for an energy of a few hundreds of MeV, while linac beams exhibit typically 1 mm transverse size, 1 $\mu rad$ divergence with 1 mm longitudinal size and $0.01~\%$ energy spread, plasma beams more likely provide 1 $\mu m$ transverse size, 1 mrad divergence with 1  $\mu m$  longitudinal size and $1~\%$ energy spread. 
Combined to the initial divergence, the energy spread can lead to significant emittance growth \cite{FloetmannPhysRevSTAB.6.034202, migliorati2013intrinsic, antici2012laser}. Collective effects and coherent synchrotron radiation can also play a role \cite{khojoyan2016transport}. 
The present LWFA electron beam properties are not directly suited for enabling FEL amplification, and electron beam manipulation is required.

\subsection{Handling of the LPA divergence}

The beam divergence requires a strong focusing.

With conventional accelerator techniques, the usually required quadrupole strength often excludes the use of electromagnetic quadruoles. Permanent magnet quadrupoles, located close to the electron source are more widely used. For example, to so-called developed QUAPEVA \cite{BenabWO2016034490, BenadWOBL14SSOQUA, marteau2017variable}, made of two quadrupoles superimposed are capable of generating a gradient of 200 T/m. The first quadrupole consists of magnets shaped as a ring and attaining a constant gradient of 155 T/m, and the second one made of four cylindrical magnets surrounding the ring and capable of rotating around their axis to achieve a gradient tunability of $\pm46 T/m$. Each tuning magnet is connected to a motor and controlled independently, enabling the gradient to be tuned with a rather good magnetic center stability ($\pm10 \mu m$) and without any field asymmetry. They are installed on translation stages, allowing the magnetic center to be adjusted. 

The focusing can also be done with a plasma itself, with a plasma lens \cite{lehe2014laser, thaury2015demonstration}, active plasma lensing \cite{Tilborg:PRL2015} or a transient magnetised plasma \cite{nakanii2015transient}. Plasma lens provides a radially symetrical focusing.

\subsection{Handling of the LPA energy spread}

A first approach consists in passing the electron beam through a demixing chicane, which sorts them in energy and reduces typically the slice energy spread by a factor of 10 \cite{Maier12, seggebrock2013bunch, couprie2014towards}. Taking advantage of the introduced correlation between the energy and the position, the slices can be focused in synchronization with the optical wave advance, in the so-called supermatching scheme \cite{loulergue2015beam}. The chicane scheme also enables to lengthen the electron bunch, for it not to escape the electron bunch because of the slippage.

A second approach to handle the large energy spread consists in using a Transverse Gradient Undulator (TGU) \cite{smith1979reducing, kroll1981theory} as considered in the early FEL days. The concept has been applied to the case of LPA \cite{Huang12, baxevanis20143d, baxevanis2015eigenmode, schroeder2013free}. The transverse gradient undulator presents usually canted magnetic poles, that generates a linear transverse dependence of the vertical undulator field in the form of $K(x)=K_{o}(1+\alpha x)$ with $\alpha$ the gradient coefficient. Associated to an optics with dispersion introducing a transverse displacement x with the energy according to $x= \eta \Delta \gamma/\gamma$, the resonant condition can be fulfilled provided $\eta=(2 + K_{o}^{2})/\alpha K_{o}^{2}$. This technique reduces the sensitivity of the FEL gain length dependence on the energy spread.

\subsection{Test experiments}

\begin{figure}[!h]
\centering\includegraphics[width=.9\linewidth]{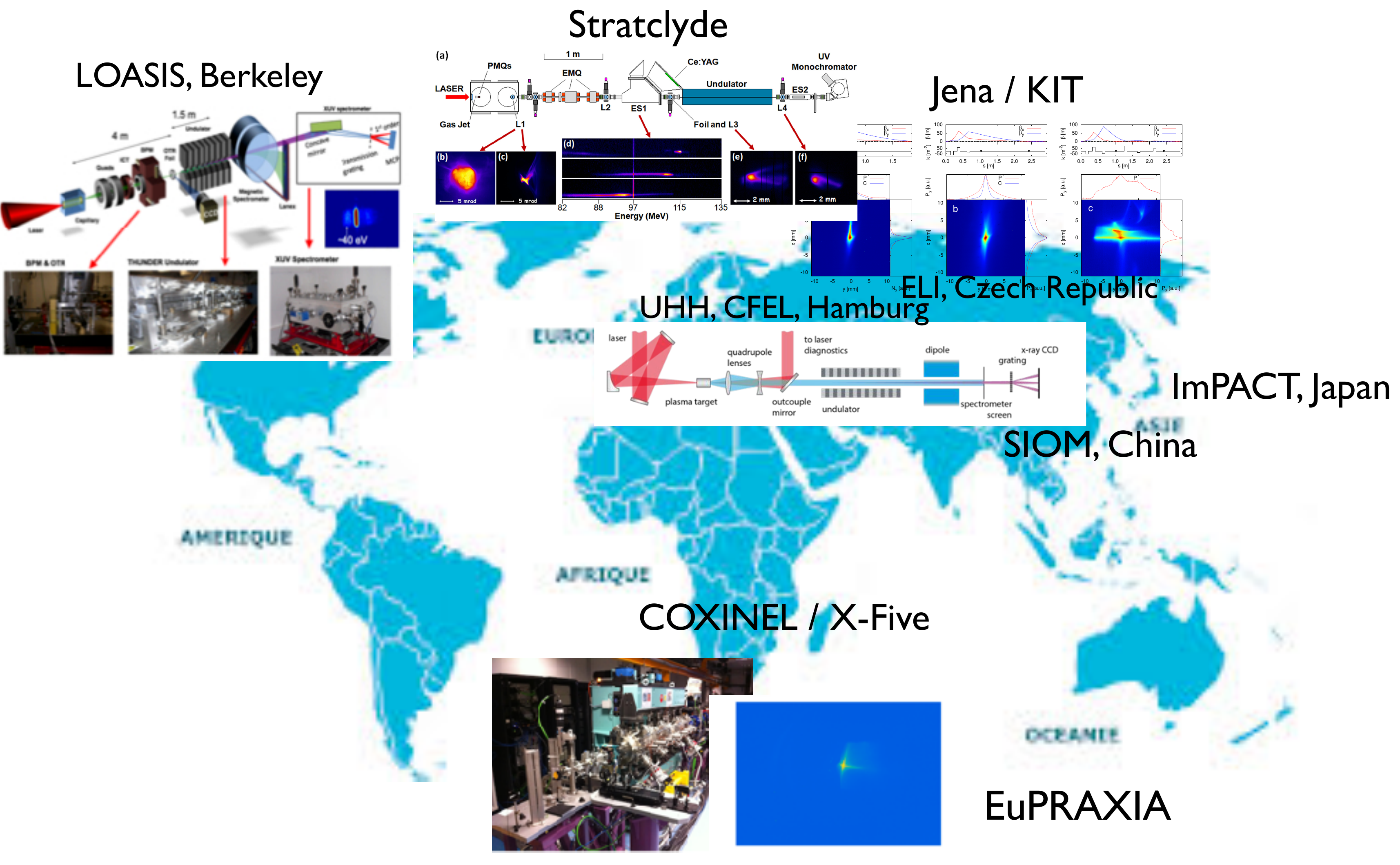}
\caption{Test experiments around the world}
\label{fig:Fig_LPAFEL}
\end{figure}

Several experiments (see Fig.~\ref{fig:Fig_LPAFEL}) are under way.

\begin{figure}[!h]
\centering\includegraphics[width=.7\linewidth]{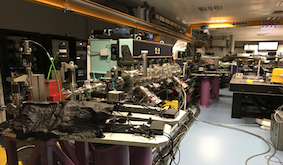}
\caption{Picture of the COXINEL experiment}
\label{fig:Fig_COX.png}
\end{figure}

The COXINEL (SOLEIL, LOA, PhLAM, France) \cite{couprie2016application, andre2016first, couprie2017coxinel} project, part of the LUNEX5 one \cite{couprie2013lunex5, couprie2014lunex5} aims at FEL amplification at 200 nm at typically 180 MeV, before increasing the energy up to 400 MeV for radiation down to 40 nm. Electrons are generated by the "salle Jaune" 2x60 TW laser in ionization configuration (see Fig.~\ref{fig:Fig_COX.png}) . Strong focusing variable strength permanent magnet quadrupoles located very close to the electron generation source handles the large electron beam divergence. A energy de-mixing chicane then deviates the electrons by 32 mm in horizontal, sorting them out in energy. The electron bunch duration is lengthened, for the photons not to escape from the electron beam distribution because of the slippage (delay between photons and electrons). A second set of quadrupoles located in front of the undulator (2 m long in-vacuum U20 or U18, then 3 m long cryo-ready undulator \cite{benabderrahmane2017development}) permits to perform a chromatic matching in the FEL interaction region, with a proper setting of the chicane. Each slice can be focused in synchronisation with the optical wave progress along the undulator. Simulations show an increase of FEL power in the supermatching condition \cite{loulergue2015beam}. The electron beam has been properly transported along the line thanks to a specific beam pointing alignment compensation enabling the separate compensation of position an dispersion. Undulator radiation has been observed.

The set-up at LBNL (USA) consists of electrons produced by a 100 TW laser in a gas jet, an active plasma lens \cite{Tilborg:PRL2015}, a chicane, a triplet, the THUNDER undulator and the magnetic beam dump. A stable jet-blade has been developed \cite{swanson2017control, barber2017measured}. 
At LUX (DESY / MPG / Univ. Hamburg), a 200 TW laser produces the electrons since 2016, and 9 nm undulator radiation has been measured in 2017 \cite{MaierCFEL}. The scheme for the FEL considers a demixing chicane. In the frame of the ImPACT collaboration in Japan, efforts are conducted to reduce the emittance and the energy spread, the pointing stability, with a very short undulator period (4 mm) for 0.4 T peak field. 

LPA based FEL experimented using the transverse gradient undulator (TGU) are implemented in a F. Shiller Univ., Jena/KIT collaboration using the JETI-40 laser,  focused in a 3 mm gas cell, an achromatic transport line, a superconducting TGU \cite{widmann2015first} and in Shanghai \cite{PhysRevAccelBeams.20.020701} with a 200 TW laser.

\section{Conclusion}

Among the panorama of light sources \cite{couprie2008xraysCRPhysique, couprie2014newgenerationJESRP, couprie2015panorama}, the advent of X-ray Free Electron Laser implemented on conventional linear accelerator took place nearly 40 years after the FEL invention, thanks to the technological developments made for colliders and step by step progresses in the FEL domain.  In parallel, the spectacular development of laser plasma acceleration (LPA) with several GeV beam acceleration in an extremely short distance appears very promising. As a first step, the qualification of the LPA with a FEL application sets a first challenge. Still, energy spread and beam divergence do not meet the stat-of-the-art performance of the conventional accelerators and have to be manipulated to fulfill the FEL requirement. Several intermediate results are very encouraging in the path towards LPA based FEL. Indeed, undulator radiation (spontaneous emission) has been seen with a simple first focusing and after transport, at 5 Hz, down to 9 nm (very short photon duration, not very intense) at LUX. The electron beam properties through a transport line, including alignment are controlled at COXINEL. «On paper» solutions for FEL amplification with typical LPA beam parameters exist with 1 $\pi$ mm.mrad, 1 $\mu m$, 1 mrad, 1 \% energy spread beam properties. In parallel, improvements of LPA performance are under way, with for example, further LPA characterization and control leading to 3.5 pC/ MeV, few percent energy spread electron beams. A design study is carried out at an European level with the EuPRAXIA collaboration \cite{walker2017horizon}. 
FEL amplification remains very challenging and constitutes an real full scale example of a demanding LPA application. Besides, sensitivity to parameters has also to be studied in depth: deviations from the optimum parameters can make the amplification no more possible;  shot to shot variations on the electron parameters and day to day reproducibility could be very critical for setting an optimum situation for attempting amplification.

\section{Acknowledgments}
This work was supported by the European Research Council under Grant COXINEL (number 340015, PI M. E. Couprie); the EuPRAXIA European Design study (653782), and the Foundation de la Coop\'eration Scientifique for the QUAPEVA contract  (2012-058T).

\section{References}
%\label{}




  \bibliographystyle{elsarticle-num} 
 \bibliography{RefCouprie.bib}


%\begin{thebibliography}{00}


%\bibitem{}

%\end{thebibliography}
\end{document}